\documentclass[twocolumn]{aastex631}

\usepackage{lineno}
\usepackage{txfonts} 
\usepackage{ulem}
\usepackage{xcolor}
\usepackage{hyperref}

\newcommand{\teff}{\mbox{T$_{\rm eff}$}}
\newcommand{\logg}{\mbox{log~{\it g}}}

\newcommand{\x}{\mbox{$\Delta_{\tiny{\mathrm{F606W,F814W}}}$}}
\newcommand{\y}{\mbox{$\Delta_{\tiny{C~\mathrm{ F606W,F814W,F322W2}}}$}}

\shorttitle{A $JWST$ project on 47\,Tucanae} 
\shortauthors{A.\,F. Marino, et al.}

\begin{document}

\title{A $JWST$ project on 47\,Tucanae.  NIRSpec spectroscopy of multiple populations among M dwarfs.
\footnote{
The JWST data presented in this article were obtained from the Mikulski Archive for Space Telescopes (MAST) at the Space Telescope Science Institute. The specific observations analyzed can be accessed via \dataset[doi: 10.17909/6t82-4360]{https://doi.org/10.17909/6t82-4360}.}
}
\author{A.\ F.\,Marino} 
\affiliation{Istituto Nazionale di Astrofisica - Osservatorio Astronomico di Padova, Vicolo dell'Osservatorio 5, Padova, IT-35122} 
\author{A.\ P.\,Milone}
\affiliation{Dipartimento di Fisica e Astronomia ``Galileo Galilei'' - Univ. di Padova, Vicolo dell'Osservatorio 3, Padova, IT-35122}
\affiliation{Istituto Nazionale di Astrofisica - Osservatorio Astronomico di Padova, Vicolo dell'Osservatorio 5, Padova, IT-35122} 
\author{A.\,Renzini}
\affiliation{Istituto Nazionale di Astrofisica - Osservatorio Astronomico di Padova, Vicolo dell'Osservatorio 5, Padova, IT-35122} 
\author{E.\,Dondoglio}
\affiliation{Istituto Nazionale di Astrofisica - Osservatorio Astronomico di Padova, Vicolo dell'Osservatorio 5, Padova, IT-35122} 
\author{E.\,Bortolan}
\affiliation{Dipartimento di Fisica e Astronomia ``Galileo Galilei'' - Univ. di Padova, Vicolo dell'Osservatorio 3, Padova, IT-35122}
\author{M.\,G.\,Carlos}
\affiliation{Theoretical Astrophysics, Department of Physics and Astronomy, Uppsala University, Box 516, SE-751 20 Uppsala, Sweden}
\author{G.\,Cordoni}
\affiliation{Research School of Astronomy and Astrophysics, Australian National University, Canberra, ACT 2611, Australia}
\author{A.\,Dotter}
\affiliation{Department of Physics and Astronomy, Dartmouth College, 6127 Wilder Laboratory, Hanover, NH 03755, USA}
\author{S.\,Jang}
\affiliation{Center for Galaxy Evolution Research and Department of Astronomy, Yonsei University, Seoul 03722, Korea}
\author{E.P.\,Lagioia}
\affiliation{South-Western Institute for Astronomy Research, Yunnan University, Kunming, 650500 P.R.China}
\author{M.\ V.\,Legnardi}
\affiliation{Dipartimento di Fisica e Astronomia ``Galileo Galilei'' - Univ. di Padova, Vicolo dell'Osservatorio 3, Padova, IT-35122}
\author{F.\,Muratore}
\affiliation{Dipartimento di Fisica e Astronomia ``Galileo Galilei'' - Univ. di Padova, Vicolo dell'Osservatorio 3, Padova, IT-35122}
\author{A.\,Mohandasan}
\affiliation{Dipartimento di Fisica e Astronomia ``Galileo Galilei'' - Univ. di Padova, Vicolo dell'Osservatorio 3, Padova, IT-35122}
\author{M.\,Tailo}
\affiliation{Istituto Nazionale di Astrofisica - Osservatorio Astronomico di Padova, Vicolo dell'Osservatorio 5, Padova, IT-35122} 
\author{T.\,Ziliotto}
\affiliation{Dipartimento di Fisica e Astronomia ``Galileo Galilei'' - Univ. di Padova, Vicolo dell'Osservatorio 3, Padova, IT-35122}

\correspondingauthor{A.\ F.\,Marino}
\email{anna.marino@inaf.it}

\begin{abstract}
We present the first spectroscopic estimates of the chemical abundance
of M dwarf stars in a globular cluster (GC), namely 47~Tucanae. By exploiting
NIRSpec on board the James Webb Space Telescope ($JWST$) we gathered
low-resolution spectra for 28 stars with masses in the range
$\sim$0.4-0.5~M$_{\odot}$. The spectra  are strongly affected by the
H$_{2}$O  water vapour bands which can be used as indicators of the
oxygen abundance. 
The spectral analysis reveals that the target stars feature a
different O abundance, with a difference of $\sim$0.40~dex between
first and the most-polluted second population. The observed range is
similar to that observed among red giant stars. 
This result reinforces previous findings based on the analysis of
photometric diagrams, including the ``chromosome maps'', providing a
first, and more direct, evidence of light element variations in the M
dwarfs' mass regime. The observation that the multiple populations,
with their variations in light elements, exhibit the same patterns
from the lower main sequence all the way to the red giant branch
further strengthens the notion that multiple stellar populations in
globular clusters formed in a series of bursts of star formation. 
\end{abstract}


\section{Introduction}\label{sec:intro}

As vividly illustrated by data from the Hubble Space Telescope
($HST$), GCs harbor multiple sequences along the color-magnitude
diagram (CMD), from below the main sequence turnoff all the way to the
red giant branch (RGB) and the asymptotic giant branch (AGB)
\citep[e.g.\,][]{piotto15, milone2015,milone2017a}. Multiband $HST$
imaging from the UV to optical has been instrumental in producing a
detailed taxonomy of the multiple population phenomenon in GCs, thanks
to different OH, NH and CN molecular blanketing due to different
abundances of these element in the various populations in each
cluster. Direct spectroscopy of photometrically-selected stars
belonging to different population has further validated the
interpretation of the photometric classification \citep{marino2019a}. 

This exquisite dataset has allowed us to constrain the detailed
properties of the multiple population phenomenon. In brief, we know
that at least two populations (first, 1P, and second, 2P, population)
can be distinguished in all the analysed GCs, with several of them
showing multiple 2P components. Thus, 1P stars have chemical
abundances similar to Halo field stars of the same metallicity
([Fe/H]), 2P stars are instead O- and C-depleted to various degrees,
and enhanced in the products of the  H-burning at high temperatures,
such as He, N, Na, Al \citep[e.g.\,][]{gratton04, marino2019a}. 

The multiple population phenomenon has been interpreted in terms of
successive stellar generations, i.e. multiple bursts of star formation
taking place within the forming GC while the chemical composition of
its ISM was (rapidly) evolving as a consequence of pollution from
evolved 1P stars. In this scenario, the observed predominance of the
2P component, contributing more than half the mass of today GCs
\citep{milone2017a}, is a challenge, as only a small fraction of the
initial 1P mass is delivered with the composition required to form 2P
stars. This is known as the ``mass budget problem''. A way out from
this is to postulate that the GC progenitors were substantially more
massive and that have lost at least 80-90\% of their 1P mass into the
Halo before delivering the naked present-day GCs  
\citep{decressin2007a, dercole08, ventura09, demink2009a, denissenkov2014a}.
One way of achieving this has been recently proposed in which the host
(dwarf) galaxy harbors extended star formation around the nascent GC,
hence the original 1P stars are distributed over a much larger volume
than the GC itself \citep{renzini2022a}. Subsequent evolution and
tidal interaction will then result in removing most of the original 1P
stars, as indicated by N-body simulations \citep{lacchin2024}. 

An alternative scenario assumes that stellar populations are
  coeval and the chemical variations are due to mass accreted onto
  existing low-mass stars, rather than being the product of different
  episodes of star formation \citep{bastian2013a}. A fascinating
  development of this scenario has been proposed by
  \citet{gieles2018a}, who suggested that 2P stars form by accretion
  of the material from extremely massive stars which can form in the
  dense environment of a proto-GC. In this scenario, there is no need
  for proto-GCs to be significantly more massive than today.

Constraining the original mass of GCs is crucial to assess what was
the contribution of GC progenitors to the assembly of the Milky Way
Halo. Moreover, depending on their mass, these stellar systems might
have substantially  contributed to  the cosmic reionization
\citep{katz13,renzini17}. All this, however, remains to be
quantitatively assessed, with high redshift observations of nascent
GCs starting to provide direct critical evidence on the formation
process \citep{vanzella16,vanzella20,adamo24}. 

Due to the observational limits imposed by the instrumentation, our
knowledge about the chemical abundance pattern of multiple populations
was, for long, confined to the most luminous phases of the
colour-magnitude diagram, e.g. the red giant stars. As most of the
stars are in fainter evolutionary stages, in the very low mass
regimes, our current knowledge on the chemical composition of stars in
GCs represents just the tip of the iceberg. Indeed, a way to
disentangle between the available scenarios is offered by the
comparison of their properties in stars with different mass. Any
accretion phenomenon should result in fact in a less efficient
capability of low mass, fully convective, stars to retain material
\citep{bondi1944a}, which translates in less pronounced chemical
variations with respect to those present in higher mass stars. Though
GCs stars have low masses, still, a sizeable range can be observed
going from very low mass stars approaching the H burning limit up to
$\sim$0.9 $\rm {M~_{\odot}}$.  

Very first investigations of multiple populations among stars with masses
down to $\sim$0.1~$\rm {M_{\odot}}$, have been conducted with near-IR
data from $HST$ WFC3 for some GCs, including NGC~2808
\citep{milone2012c}, NGC~6121 \citep[M~4,][]{milone2014} and others
\citep{milone2017b,milone2019}, and a  survey for several clusters was
recently presented by \citet{dondoglio2022a}. 
These analyses, entirely based on photometric diagrams combined with
synthetic spectra, indicate that internal variations in chemical
abundances persist all the way to very low mass stars, and these
variations are quantitatively similar to those observed among the
giants in the same cluster. In practice, chemical differentiation is
possible thanks to blanketing in the F160W band due to water vapour,
thus tracing differences in oxygen abundance. Thus, multiple lower
main sequences have been identified which appear to come in the same
proportion of the multiple sequences already identified among e.g.,
the RGB stars of the same clusters. In \citet{dondoglio2022a} the
present-day mass function in the explored stellar mass range was
estimated for the various 1P and 2P populations, finding to be
consistently the same (see also \citealt{milone2019, scalco24}).

Now, thanks to the new facilities on board the James Webb Space
Telescope ($JWST$), the observation of GC stars down to very faint
magnitudes has become considerably more efficient than with previous
telescopes. 
First evidence of multiple stellar populations in very low
mass stars below the main sequence knee have been reported for
NGC\,104 \citep[47~Tucanae,][]{milone2023a},  M\,92
\citep{ziliotto2023a}, and NGC\,6440 \citep{cadelano2023}, using
H$_2$O blaketing (e.g.\, in the F150W, F200W or F322W2 bands). 
Most recently, the CMD of 47~Tucanae from very deep $JWST$/NIRCam data
has reached, for the first time, the H-burning limit and the more
luminous brown dwarfs \citep{marino2024}. 

In this work, we take advantage of the capabilities of the $JWST$
multi-object spectrograph NIRSpec to characterise  the multiple
stellar populations among the M dwarfs of this cluster.  
Until now, the spectroscopic observation of such faint stars was
prohibitive, but $JWST$ with NIRSpec  allows us to make a big step
forward in this context. 
We in fact have reached the faintest stars ever observed in a GC and
use these data to document the multiple stellar population phenomenon
in M dwarfs from {\it direct} spectroscopic measurements, thus
providing a consistent characterisation combing photometric and
spectroscopic data. 

The layout of this work is as follows: Section~\ref{sec:data} presents
the observations, Section~\ref{sec:analysis} the analysis, and finally
Section~\ref{sec:results} discusses the results.

\section{The dataset}\label{sec:data}

We have acquired $JWST$ data for 47~Tucanae  from parallel
observations including both spectra from NIRSpec and photometric data
from NIRCam (GO2560, PI Marino).
NIRSpec data were gathered by using the G235M/F170LP disperser-filter
combination which provides a 1.66-3.07~$\mu$m range in wavelength and
a resolution of $R\sim$1000.  
To simultaneously observe many targets, we employed the multi-object
spectroscopy (MOS) mode using the micro-shutter assembly (MSA)
configuration. 
In parallel, NIRCam observed in the two filters $F115W$ and $F322W2$,
those that most efficiently ensure the separation of multiple
sequences among M dwarfs \citep{milone2023a, marino2024}, again using
H$_2$O blanketing but this time in the F322W2 passband. 

Observations were gathered in July 2022 in the first $JWST$ cycle, but
they experienced technical issues. Specifically, wind-tilt events
affected the quality of the data \citep[see][for more
details]{milone2023a}; and nodding issues resulted in missing spectra
in one of the two nod positions. In the end, while photometric images
were analysed anyway in \citet{milone2023a}, NIRSpec data were not
suited for a straightforward analysis, with only a few stars having
spectra of sufficient quality. 

New observations were then completed in September 2023. We employed
the same strategy of the first observations and gathered spectroscopic
and photometric data in parallel. This time we did not experience any
failure, and obtained new images and spectra. While for the
photometric analysis and the presentation of the first spectra we
refer to \citet{marino2024}, here we analyse the full spectroscopic
dataset. 
 
NIRSpec spectra for a sample of 29 M dwarfs were obtained in the
magnitude interval $19.5\lesssim m_{\rm F160W} \lesssim 20.3$,
corresponding to stellar masses between 0.35 and 0.5 $M_{\odot}$. 
Our exposures were organised in 16 groups, and the NRSIRS2 was
employed as readout pattern, with a total on-target time of 47,280s
for each of the observed stars. 
 
The exposures were sub-divided into 2-Shutter nodding positions, and a
dither pattern of 100 shutters has been applied, which corresponds to
20$\arcsec$ in the dispersion direction. 
Each exposure was integrated for 1,182s. 
Of our 28 M dwarf targets, 17 are observed in both dither positions,
resulting in a total of 40 exposures, while the remaining 12 only in
one dither position providing half of the total exposure. 
The 1-D extracted spectra processed with the $JWST$ Science
Calibration Pipeline \citep{Bushouse2023} have been used for the
analysis.  

The location of the NIRSpec and NIRCam parallel fields are shown in
\cite{marino2024} (see their Figure~1).
Specifically, the NIRSpec field has pointed the same field  previously
covered with $HST$, for which the filters F110W and F160W from GO11453
are available. The spectroscopic targets have been selected 
to sample the entire range in color as observed from the $m_{\rm
  F160W}$ vs. $m_{\rm F110W} - m_{\rm F160W}$ CMD that we obtained
from $HST$ archive data. This CMD-based selection has ensured a proper
sampling of all the stellar populations hosted in the cluster. 

The signal-to-noise ratio of our spectra ranges from $\sim$60 to
$\sim$35 per pixel mostly depending on whether the target has been
observed in both or in only one of the nod positions, respectively. 
An example of our spectra is shown in Figure~\ref{fig:obs}, for the
two stars \#32620 and \#25862, with very similar atmospheric
parameters, log$g$ and $T_{\rm eff}$. These spectra will be further
discussed below, but we can already note that \#32620  has much
stronger H$_2$O molecular bands than the other.  Note that this
spectrum has a gap, due to the physical edges of the 2 detectors in
the MSA, that depends on the shutter position and the disperser. Some
of the gathered spectra were not analyzed because they are affected by
wavelength cutoffs in crucial spectral regions for our analysis. In
the spectrum of \#32620 the gap is in a region not affected by the
H$_{2}$O molecules, which does not prevent the analysis of this star. 

\section{Spectral analysis}\label{sec:analysis}

\begin{figure*}
\centering
	\includegraphics[width=1.0\textwidth]{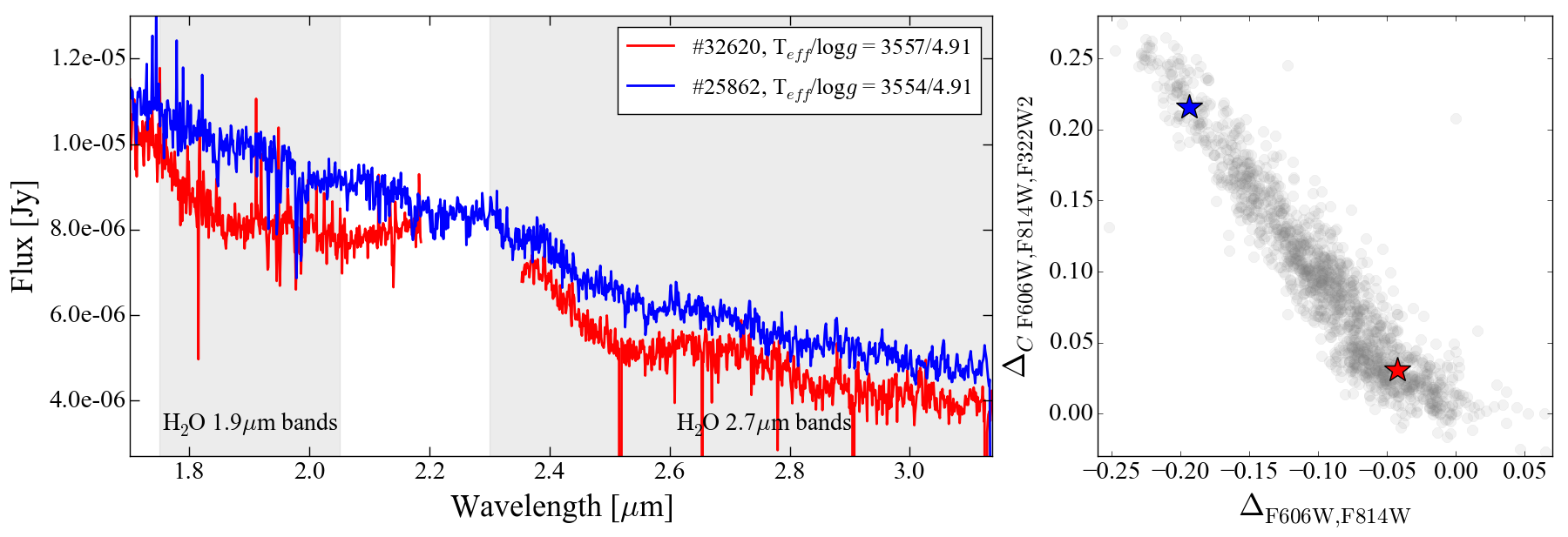} 
    \caption{NIRSpec spectra for the two stars \#32620 and \#25862 marked with large starred symbols in the ChM from \citet{marino2024} (right panel). Based on the location on the ChM, the two stars have different values both in \x\, and \y, with \#32620 being a 1P star, and \#25862 a 2P one. The similar values of the atmospheric parameters for the plotted stars have been reported in the inset of the left panel. The shaded regions indicate the spectral range spanned by the H$_{2}$O bands.}
    \label{fig:obs}
\end{figure*}

A visual comparison of the two spectra shown in Fig.~\ref{fig:obs}
\citep[see also Fig.~8 of][for another pair of stars]{marino2024}
clearly reveals   different absorption for stars belonging to
distinct stellar populations. The position of these two stars on the
cluster {\it chromosome map} (ChM, see \citealt{milone2017a}) shown in
the right panel of Fig.~\ref{fig:obs}, suggests that \#32620 is a 1P
star whereas \#25862 is one of the most-extremely polluted 2P stars,
which we expect to be associated with the lowest oxygen abundance. 

In the following we describe our employed technique to estimate the
difference in the chemical abundances of the observed M dwarfs. The
spectral analysis has been conducted by comparing the   observed
spectra to synthetic spectra in the range from 1.5 to 2.2~$\mu$m,
where the H$_2$O absorption is deepest.  

\subsection{Synthetic spectra}\label{sec:synthetic}

The theoretical spectra have been computed by using ATLAS and SYNTHE
programs{\footnote{http://kurucz.harvard.edu}} and $\alpha$-enhanced
model atmospheres with \teff\ ranging from 3500 to 4000~K, and \logg\
from 4.5 to 5.0, while metallicity has been set to [Fe/H]=$-0.75$~dex
\citep[e.g.\,][]{carretta2009b, marino2023a}. The abundances of C, N,
and O, have been varied to mimic the chemical diversity between stars
belonging to different stellar populations in GCs{\footnote{Other
    elements, such as Na and Al, are not expected to produce changes
    in the overall NIRSpec spectra.}}.  
Spectral synthesis includes all the lines and molecules listed in the
Kurucz website, including the H$_2$O molecular bands
\citep{partridge1997a} that dominate the NIR spectra of M dwarfs. 

\begin{figure*}
	\includegraphics[width=1.\textwidth]{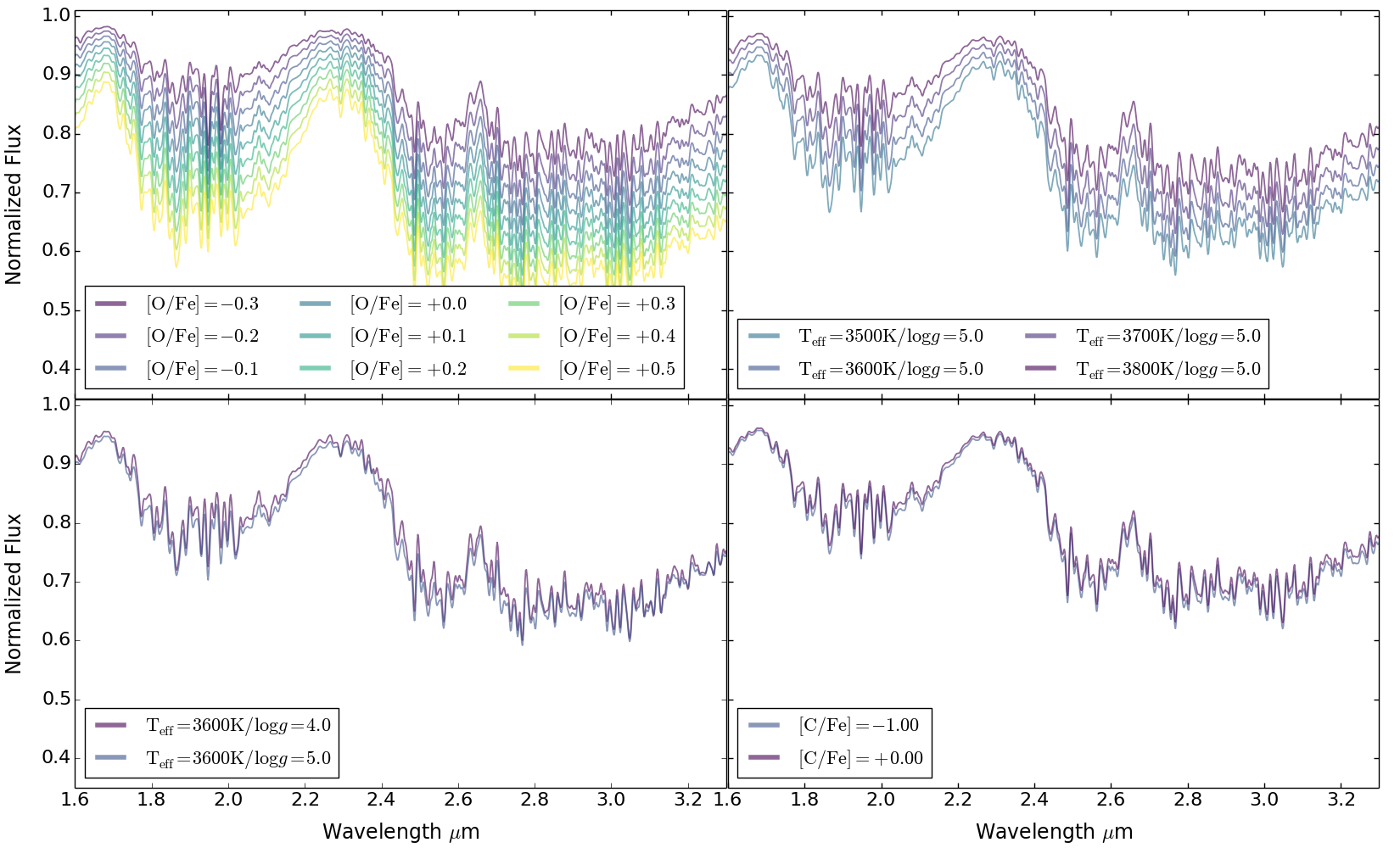} 
    \caption{Synthetic spectra computed at different O abundances, keeping all the other chemical abundances and stellar parameters fixed, are shown in the top-left panel. As indicated in the inset, the simulations have been computed for [O/Fe] ranging from $-$0.3 to $+$0.5~dex, with a step of 0.10~dex. The other panels illustrate the impact of changing temperatures (top-right) up to 300~K (in steps of 100~K), \logg\ by 1~dex
    (bottom-left), and C abundances (bottom-right) by 1~dex.}
    \label{modelsTeff}
\end{figure*}

Figure~\ref{modelsTeff} displays examples of the synthetic spectra
with changing one at a time oxygen, effective temperatures (\teff),
surface gravities (\logg), and carbon. An inspection at these
theoretical spectra immediately indicate that oxygen abundances
heavily impact on the shape of the spectra, with O-enhanced stars
highly absorbed all over the observed range in wavelength (top-left
panel). Besides O, the next element affecting the spectra is carbon
(bottom-right panel), which subtracts oxygen to H$_2$O by locking it
into CO. However, increasing [C/Fe] by 1~dex, from $-$1.00 to $+$0.00,
a range often observed among different stellar populations in GCs,
only marginally diminishes the overall water absorption. The impact of
the other elements that are observed to vary among higher mass stars
in GCs, e.g. N and Mg, is negligible in the NIR wavelength range
explored here. 

Together with O abundances, the effective temperature is the parameter
that more heavily affects the spectra. Decreasing \teff\ by 100~K
produces a similar effect as increasing O by ~0.10~dex. Instead, the
effect of surface gravity is negligible. 

This discussion highlights the sensitivity to O of our NIRSpec spectra
of M dwarfs, which is one of the main chemical players in the multiple
stellar population phenomenon. Once the \teff\ is fixed, as discussed
in next Section, we can  proceed  to estimate the O abundances from
the spectra of these M dwarfs. 

\subsection{Observed spectra}\label{sec:observed}

The observed spectra have been compared to synthetic ones with the
\teff/\logg\ values obtained from the position of each star on the
CMD. Specifically, the $m_{\rm F814W}$ vs. $m_{\rm F606W}-m_{\rm
  F814W}$ CMD has been fitted with $\alpha$-enhanced isochrones at a
metallicity [Fe/H]=$-$0.7~dex from the Dartmouth database
\citep{dotter2008a}. Distance modulus has been fixed to 13.23~mag,
reddening and age to $E(B-V)=$0.02~mag and 12.5~Gyr, respectively.
Then, we assume as \teff/\logg\ for each star the values of the
isochrone at the same mag level. A list of the adopted parameters,
together with the estimated stellar mass, is
presented in Table~\ref{tab:targets}. 
While with this approach our estimates can be affected by systematics,
we are interested here in the relative star-to-star abundance
difference to infer the overall O range in these low mass stars. 

We notice here that the lower mass in our target sample is 0.35~${\rm M_{\odot}}$, with a total of
  twelve stars with mass smaller than 0.40~${\rm M_{\odot}}$,
  the approximate mass below which stars are fully convective.
Hence, these stars are potentially the most important benchmarks for
constraining the formation scenario of multiple stellar populations.

To compare  with synthetic models, we need to normalise the observed
spectra. Although the true-continuum level cannot be seen in the
observed spectrum, we can exploit a pseudo-continuum. We take
advantage of the two spectral regions around 1.7 and 2.3 $\mu$m, not
heavily absorbed by H$_{2}$O molecules. 
These two regions define the edges of the spectral range used for the
synthesis. The same pseudo-continuum used for the observed spectra has
been assumed for the synthetic spectra. Then, we perform a $\chi^{2}$
minimisation to find the best-fit model. 
In this way the region exploited to estimate O abundances includes the
H$_{2}$O set of bands centered at 1.9~$\mu$m, which span the spectral
range from 1.75 to 2.05 $\mu$m (see Fig.~\ref{fig:obs}). For the
second band visible in our data, e.g.\, the H$_{2}$O band around 2.7
$\mu$m, a pseudo-continuum is much harder to define, and we prefer not
to use this part of the spectrum. 

\begin{figure*}
	\includegraphics[width=1.\textwidth]{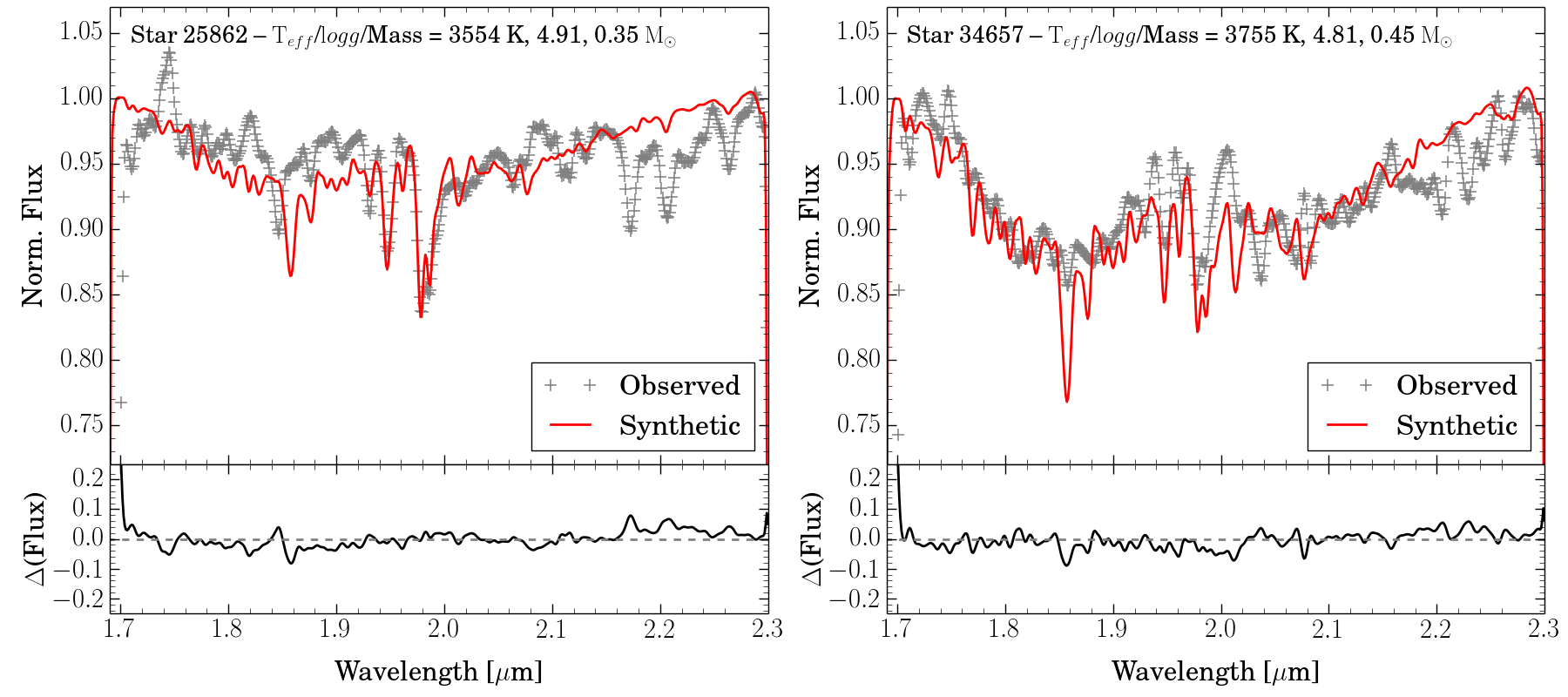} 
    \caption{Examples of two spectral synthesis. The displayed observed spectra (grey crosses) are relative to stars \#25862 and \#34657 which have different locations on the chromosome map. Overall, the analysed spectral region of star \#34657, photometrically classified as a 1P star, are deeper than those of \#25862, which is associated with the most-polluted 2P population. This is in spite of the lower temperature of the latter which acts in the direction of deeper H$_{2}$O features (see Fig.~\ref{modelsTeff}). The best-fit synthetic spectrum is shown in red. Both the observed and the simulated spectra have been normalized at the same pseudo-continuum, and the difference of the simulated$-$observed ($\Delta$(Flux)) shown in the bottom panels.} 
    \label{fig:sintesi}
\end{figure*}

Figure~\ref{fig:sintesi} shows some examples of the comparison between
predicted and observed spectra normalised at the same
pseudo-continuum. Even if the synthetic spectra do not reproduce all
the observed features, probably due to the used line lists missing
some features, still the overall spectral pattern is satisfactorily
reproduced.  
Specifically, the right panel illustrates a match for a 1P star, with
the best-fit synthetic spectrum having [O/Fe]$=+$0.32~dex, while for
the spectrum on the right, which belongs to a 2P star, the abundance
that best matches the observations is [O/Fe]$=-$0.25~dex. The
spectrum of the O-enhanced star associated with the 1P displays much
deeper H$_{2}$O  band features (note that this pattern is clear even
if this spectrum belongs to a
star with a temperature higher than the one of the displayed 2P star,
which reduces the difference due to oxygen abundances, as discussed in
the previous section).
The oxygen abundances inferred for the 28 analyzed M dwarfs are listed
in Table~\ref{tab:targets}. 

As clearly shown by the synthetic spectra discussed in
Section\,\ref{sec:synthetic}, our estimates are mostly internally
affected by uncertainties in temperatures, that have been fixed by
isochrones. As discussed, the photometric errors suggest
  that the internal uncertainties in temperatures is as small as
  $\lesssim$50~K which correspond to an error in [O/Fe] of
  $\sim$0.05~dex. We note however that our temperatures could be
  affected by systematics that can be much larger than the internal
  errors.
The overall quality of the spectra, including the limited S/N and the
fitting procedure, introduce an additional random error of
$\sim$0.15~dex. 

Although, in general, systematic errors 
  are expected to affect abundances uniformely, not changing the
  internal range in the chemical content, we discuss here possible
  systematics that could instead depend on the actual abundances of
  each star. Indeed, changes in the light element abundances,
  expecially in helium, do impact on the adopted boundary conditions
  used in the stellar model computations and this impacts on the
  predicted effective temperature scale \citep{pietrinferni21,
    vandenberg23}. The average internal variation in He between 2P and
  1P stars in 47~Tucanae
is $\Delta \rm{(Y)}=0.011\pm0.005$, with a maximum variation of
$\Delta \rm{(Y)}=0.049\pm0.005$ between the most extreme (in He
enhancement/O depletion) 2P and the 1P stars \citep{milone2018a}. We have verified, by using
He-enhanced isochrones, that an increase in temperature around 10~K
is associated by assuming the average enhancent in He, while if we
consider the most-extremly He enhanced population, the temperature
rises from $\sim$30 for the coolest stars up to $\sim$50~K for the
warmest ones. As previously discussed, rising temperatures systematically decreases
the inferred
O chemical content. However, the small variations at play only
marginally change the O abundances, by at most $\sim$0.05~dex.

\section{Results and Discussion}\label{sec:results}

\begin{figure*}
\includegraphics[width=.7\textwidth]{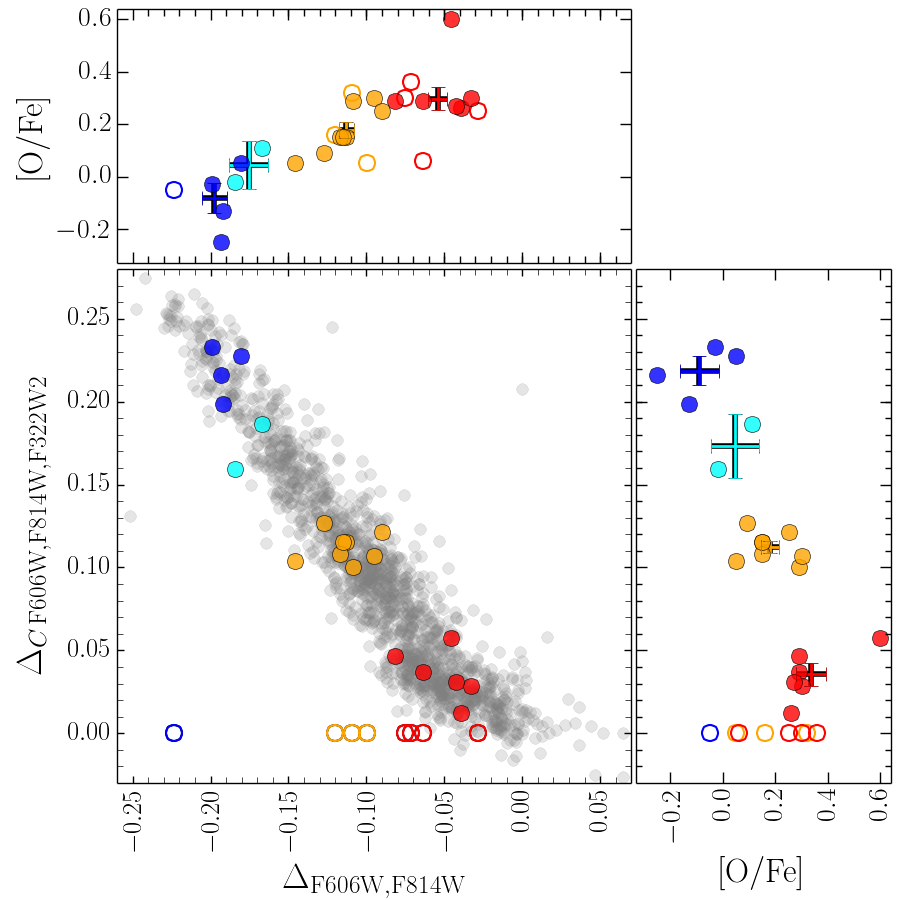} 
\centering
    \caption{ChM diagram of M dwarfs in 47~Tucanae from $JWST$-$HST$ photometry (grey dots). Spectroscopic targets observed with NIRSpec@$JWST$ are represented with filled coloured circles. Targets with no available $m_{\rm F322W2}$ from NIRCam@$JWST$ are plotted with \y=0.00 as empty circles. Different colors have been used for stars associated with different stellar populations on the ChM, specifically: first population (1P) stars are colored in red, second population ones have been colored in orange (2Pa), cyan (2Pb) and blue (2Pe). The [O/Fe] abundance vs. the \x\ and \y\ values is shown in the upper and right panels, respectively. For each population, we plot the average values and the associated error.}
    \label{fig:ChM_O}
\end{figure*}

The comparison of the spectra illustrated in previous sections
indicates the existence of a  sizable spread in oxygen abundances
among the M dwarfs of 47~Tucanae. Figure\,\ref{fig:ChM_O} shows the
location of all the analyzed targets on the ChM of the cluster,
indicating that the whole stellar-populations pattern of 47~Tucanae
has been sampled by our NIRSPec spectra.  
We divide the ChM into four  stellar populations, from the 1P
(represented in red) up to the extreme population 2Pc (represented in
blue), through two intermediate populations 2Pa and 2Pb (represented
in orange and cyan, respectively).  We can then inspect the O
abundances of the selected groups. 

From the O abundance distributions as a function of \x\ and \y\
plotted in Figure\,\ref{fig:ChM_O}, we notice a clear trend  along the
ChM sequence. Specifically, there is a correlation between [O/Fe] and
\x, and an anticorrelation with \y. The average abundances for each
population, with relative errors determined as the r.m.s.$/\sqrt{N-1}$
(with $N$ the number of stars in each group), are listed in
Table~\ref{tab:average}. From these values we have that the difference
in the average oxygen abundance between the 1P and the population 2Pa
is $\Delta(\rm {[O/Fe]_{1P}-[O/Fe]_{2Pa}})=0.12\pm0.05$~dex, which
means at a level of $\sim 2.5~\sigma$. Moving upwards in the ChM, the
mean difference with the 2Pb population increases to $\Delta(\rm
{[O/Fe]_{1P}-[O/Fe]_{2Pb}})=+0.26\pm0.10$~dex, but, given that only
two stars are analyzed in the 2Pb, the significance remains marginally
less than $3~\sigma$. The oxygen abundance is observed to more
significantly decrease in the 2Pc populations (with 5 [O/Fe] estimates
available), with $\Delta(\rm
{[O/Fe]_{1P}-[O/Fe]_{2Pe}})=0.38\pm0.07$~dex, a $>5~\sigma$
difference. Note that also the oxygen difference between 2Pa and 2Pc
is at a level of more than $3~\sigma$, e.g.\, $\Delta(\rm
{[O/Fe]_{2Pa}-[O/Fe]_{2Pe}})=0.26\pm0.07$~dex. 
We note that the r.m.s. of each group is between $\sim$0.10 and
$\sim$0.15~dex (see Table~\ref{tab:average}), which is smaller than
the internal error that is estimated to be associated with the
individual measurements (in Section~\ref{sec:observed}), which might
have been overestimated. 

Compared to the literature values available for more massive stars,
our [O/Fe] values are systematically underestimated (up to 0.15~dex), 
a pattern that can be expected given the diversity in the stellar
parameters and the different used spectral features \citep{collet2007,
  amarsi2016}. On the other hand, the goal of this analysis is to enstablish the
range of the oxygen abundances spanned by very low mass stars, instead
of the real abundances. 
In this context, we conclude that a total range of
$\Delta$[O/Fe]=0.38~dex, found in this study, is fully consistent with
the range exhibited by the red giants, as from high-resolution spectra
\citep[e.g.\,][]{carretta2009a, dobrovolskas2014a}. 

The presence of the same abundance variations in stars with different
mass, from $\sim$0.8 down to $\sim$0.4~M$_{\odot}$, suggests that, in
old Milky Way GCs, the multiple stellar populations phenomenon does 
not depend on the stellar mass. 
This fact represents a challenge for any accretion scenario
\citep[e.g.\,][]{bastian2013a, gieles2018a} where the
chemical pollution degree is expected to depend on the stellar mass. 
We notice that even if only a subsample of our target stars might
reside below the mass limit of a fully convective regime, our results
are a direct spectroscopic confirmation of what previously
photometrically observed for stars fainter than the main sequence
knee, from $\sim$0.5~${\rm M_{\odot}}$ down to $\sim$0.1~${\rm M_{\odot}}$. 
Similar abundance patterns among today GC main sequence, red giants
and low mass stars down to the H-burning limit, 
supports the scenario that the multiple stellar populations
formed in different bursts of star formation. 

We conclude by enphasising that, thanks to the multi-object
capabilities of NIRSpec combined with the high efficiency of the
$JWST$ in the infrared region of the spectrum, we have measured the
first spectroscopic chemical abundances for M dwarfs in a GC. These
observations provides the spectroscopic data of the faintest stars
ever observed in these old stellar systems. 
Our results confirm the presence of a genuine oxygen internal spread
among M dwarfs of 47~Tucanae. For the first time, the existence of a
spread in light elements chemical abundances of M dwarfs has been constrained
directly from spectra, confirming previous results based on
photometric diagrams \citep[][]{milone2012b, milone2014,
  dondoglio2022a, milone2023a, marino2024}.

\begin{table*}
\centering
\caption{Coordinates, adopted stellar temperatures and surface gravities, mass, location on the chromosome map, and inferred [O/Fe] values for the NIRSpec targets.}\label{tab:targets}
\begin{tabular}{c c c c c c c c r}
\hline\hline
    ID  &  RA & DEC & \teff& \logg &  Mass  &\x&\y  &[O/Fe]\\
        & (J2000) & (J2000) & (K) & (cgs) & (M$_{\odot}$) & (mag) & (mag) & dex\\
\hline
   3001 & 5.63973525 & $-$72.09532277 & 3602 &  4.89 &  0.38  &  $-$0.0637 &  0.0368  &    0.29  \\
   3819 & 5.64094229 & $-$72.09409725 & 3659 &  4.86 &  0.41  &  $-$0.1669 &  0.1868  &    0.11  \\
   8749 & 5.67429569 & $-$72.08765474 & 3579 &  4.90 &  0.36  &  $-$0.1084 &  0.1005  &    0.29  \\
  11213 & 5.65667045 & $-$72.08465712 & 3759 &  4.81 &  0.45  &  $-$0.0391 &  0.0123  &    0.26  \\
  12058 & 5.58636195 & $-$72.08445120 & 3904 &  4.75 &  0.49  &  $-$0.0952 &  0.1070  &    0.30  \\
  13694 & 5.70320857 & $-$72.08262859 & 3745 &  4.81 &  0.44  &  $-$0.1133 &  0.1152  &    0.15  \\
  14443 & 5.74027949 & $-$72.08185326 & 3567 &  4.91 &  0.36  &  $-$0.0899 &  0.1214  &    0.25  \\
  15654 & 5.55751693 & $-$72.07992265 & 3569 &  4.91 &  0.36  &  $-$0.1843 &  0.1596  & $-$0.02  \\
  17065 & 5.57984182 & $-$72.07817684 & 3643 &  4.87 &  0.40  &  $-$0.1270 &  0.1268  &    0.09  \\
  17470 & 5.59731765 & $-$72.07867065 & 3785 &  4.79 &  0.46  &  $-$0.1172 &  0.1084  &    0.15  \\
  19400 & 5.57128890 & $-$72.07541568 & 3736 &  4.82 &  0.44  &  $-$0.1455 &  0.1042  &    0.05  \\
  20051 & 5.72710583 & $-$72.07569270 & 3788 &  4.79 &  0.46  &  $-$0.0454 &  0.0576  &    0.60  \\
  20391 & 5.55149616 & $-$72.07522199 & 3667 &  4.85 &  0.41  &  $-$0.1917 &  0.1984  & $-$0.13  \\
  20477 & 5.56336608 & $-$72.07397661 & 3792 &  4.79 &  0.46  &  $-$0.1991 &  0.2330  & $-$0.03  \\
  22777 & 5.68245973 & $-$72.07146076 & 3668 &  4.85 &  0.41  &  $-$0.0327 &  0.0285  &    0.30  \\
  23062 & 5.70849703 & $-$72.07175055 & 3758 &  4.81 &  0.45  &  $-$0.1803 &  0.2276  &    0.05  \\
  25260 & 5.59224113 & $-$72.06959101 & 3618 &  4.88 &  0.38  &  $-$0.0818 &  0.0465  &    0.29  \\
  25862 & 5.75511193 & $-$72.06753918 & 3554 &  4.91 &  0.35  &  $-$0.1934 &  0.2159  & $-$0.25  \\
  32620 & 5.62479859 & $-$72.06054481 & 3557 &  4.91 &  0.35  &  $-$0.0423 &  0.0311  &    0.42  \\
  34657 & 5.64627470 & $-$72.05729149 & 3755 &  4.81 &  0.45  &  $-$0.0749 &    --    &    0.30  \\
  36425 & 5.64309275 & $-$72.05518242 & 3902 &  4.75 &  0.49  &  $-$0.1091 &    --    &    0.32  \\	 
  37942 & 5.66283930 & $-$72.05331521 & 3568 &  4.91 &  0.36  &  $-$0.0637 &    --    &    0.06  \\
  40504 & 5.61419155 & $-$72.04907397 & 3663 &  4.85 &  0.41  &  $-$0.1150 &  0.1152  &    0.15  \\
  40579 & 5.65405635 & $-$72.04884485 & 3713 &  4.83 &  0.43  &  $-$0.2234 &    --    & $-$0.05  \\
  41668 & 5.67021510 & $-$72.04763272 & 3575 &  4.90 &  0.36  &  $-$0.0998 &    --    &    0.05  \\
  41764 & 5.64695293 & $-$72.04810886 & 3637 &  4.87 &  0.39  &  $-$0.0282 &    --    &    0.25  \\
  42484 & 5.63151002 & $-$72.04646898 & 3821 &  4.78 &  0.47  &  $-$0.0711 &    --    &    0.36  \\
  48348 & 5.65298871 & $-$72.03644822 & 3573 &  4.90 &  0.36  &  $-$0.1199 &    --    &    0.16  \\\hline
\end{tabular}
\end{table*}

\begin{table}
\centering
\caption{Average [O/Fe] abundances, associated error, and r.m.s. of the inferred for the four stellar populations selected on the M dwarf chromosome map. The number of stars (\#) for each population is also indicated.}\label{tab:average}
\begin{tabular}{c c c c c }\hline
\small
   & [O/Fe] & $\pm$ & r.m.s.&\# \\  \hline
1P &  $+$0.30      &  0.04  & 0.13  &  10 \\
2Pa&  $+$0.18      &  0.03  & 0.09  &  11 \\
2Pb&  $+$0.04      &  0.09  & 0.09  &  2  \\
2Pc&  $-$0.08      &  0.06  & 0.13  &  5  \\\hline
\end{tabular}
\end{table}

\bibliography{../47Tuc_Marino.bbl}{}

\begin{thebibliography}{}
\expandafter\ifx\csname natexlab\endcsname\relax\def\natexlab#1{#1}\fi
\providecommand{\url}[1]{\href{#1}{#1}}
\providecommand{\dodoi}[1]{doi:~\href{http://doi.org/#1}{\nolinkurl{#1}}}
\providecommand{\doeprint}[1]{\href{http://ascl.net/#1}{\nolinkurl{http://ascl.net/#1}}}
\providecommand{\doarXiv}[1]{\href{https://arxiv.org/abs/#1}{\nolinkurl{https://arxiv.org/abs/#1}}}

\bibitem[{{Adamo} {et~al.}(2024){Adamo}, {Bradley}, {Vanzella}, {Claeyssens},
  {Welch}, {Diego}, {Mahler}, {Oguri}, {Sharon}, {Abdurro'uf}, {Hsiao},
  {Messa}, {Zackrisson}, {Brammer}, {Coe}, {Kokorev}, {Ricotti}, {Zitrin},
  {Fujimoto}, {Inoue}, {Resseguier}, {Rigby}, {Jim{\'e}nez-Teja}, {Windhorst},
  \& {Xu}}]{adamo24}
{Adamo}, A., {Bradley}, L.~D., {Vanzella}, E., {et~al.} 2024, arXiv e-prints,
  arXiv:2401.03224, \dodoi{10.48550/arXiv.2401.03224}

\bibitem[{{Amarsi} {et~al.}(2016){Amarsi}, {Asplund}, {Collet}, \&
  {Leenaarts}}]{amarsi2016}
{Amarsi}, A.~M., {Asplund}, M., {Collet}, R., \& {Leenaarts}, J. 2016, \mnras,
  455, 3735, \dodoi{10.1093/mnras/stv2608}

\bibitem[{{Bastian} {et~al.}(2013){Bastian}, {Lamers}, {de Mink}, {Longmore},
  {Goodwin}, \& {Gieles}}]{bastian2013a}
{Bastian}, N., {Lamers}, H.~J.~G.~L.~M., {de Mink}, S.~E., {et~al.} 2013,
  \mnras, 436, 2398, \dodoi{10.1093/mnras/stt1745}

\bibitem[{{Bondi} \& {Hoyle}(1944)}]{bondi1944a}
{Bondi}, H., \& {Hoyle}, F. 1944, \mnras, 104, 273,
  \dodoi{10.1093/mnras/104.5.273}

\bibitem[{{Bushouse} {et~al.}(2023){Bushouse}, {Eisenhamer}, {Dencheva},
  {Davies}, {Greenfield}, {Morrison}, {Hodge}, {Simon}, {Grumm}, {Droettboom},
  {Slavich}, {Sosey}, {Pauly}, {Miller}, {Jedrzejewski}, {Hack}, {Davis},
  {Crawford}, {Law}, {Gordon}, {Regan}, {Cara}, {MacDonald}, {Bradley},
  {Shanahan}, {Jamieson}, {Teodoro}, \& {Williams}}]{Bushouse2023}
{Bushouse}, H., {Eisenhamer}, J., {Dencheva}, N., {et~al.} 2023, {JWST
  Calibration Pipeline}, 1.10.0,  Zenodo, \dodoi{10.5281/zenodo.7795697}

\bibitem[{{Cadelano} {et~al.}(2023){Cadelano}, {Pallanca}, {Dalessandro},
  {Salaris}, {Mucciarelli}, {Leanza}, {Ferraro}, {Lanzoni}, {Rosie Chen},
  {Freire}, {Heinke}, \& {Ransom}}]{cadelano2023}
{Cadelano}, M., {Pallanca}, C., {Dalessandro}, E., {et~al.} 2023, \aap, 679,
  L13, \dodoi{10.1051/0004-6361/202347961}

\bibitem[{{Carretta} {et~al.}(2009{\natexlab{a}}){Carretta}, {Bragaglia},
  {Gratton}, {D'Orazi}, \& {Lucatello}}]{carretta2009b}
{Carretta}, E., {Bragaglia}, A., {Gratton}, R., {D'Orazi}, V., \& {Lucatello},
  S. 2009{\natexlab{a}}, \aap, 508, 695, \dodoi{10.1051/0004-6361/200913003}

\bibitem[{{Carretta} {et~al.}(2009{\natexlab{b}}){Carretta}, {Bragaglia},
  {Gratton}, \& {Lucatello}}]{carretta2009a}
{Carretta}, E., {Bragaglia}, A., {Gratton}, R., \& {Lucatello}, S.
  2009{\natexlab{b}}, \aap, 505, 139, \dodoi{10.1051/0004-6361/200912097}

\bibitem[{{Collet} {et~al.}(2007){Collet}, {Asplund}, \&
  {Trampedach}}]{collet2007}
{Collet}, R., {Asplund}, M., \& {Trampedach}, R. 2007, \aap, 469, 687,
  \dodoi{10.1051/0004-6361:20066321}

\bibitem[{{de Mink} {et~al.}(2009){de Mink}, {Pols}, {Langer}, \&
  {Izzard}}]{demink2009a}
{de Mink}, S.~E., {Pols}, O.~R., {Langer}, N., \& {Izzard}, R.~G. 2009, \aap,
  507, L1, \dodoi{10.1051/0004-6361/200913205}

\bibitem[{{Decressin} {et~al.}(2007){Decressin}, {Meynet}, {Charbonnel},
  {Prantzos}, \& {Ekstr{\"o}m}}]{decressin2007a}
{Decressin}, T., {Meynet}, G., {Charbonnel}, C., {Prantzos}, N., \&
  {Ekstr{\"o}m}, S. 2007, \aap, 464, 1029, \dodoi{10.1051/0004-6361:20066013}

\bibitem[{{Denissenkov} \& {Hartwick}(2014)}]{denissenkov2014a}
{Denissenkov}, P.~A., \& {Hartwick}, F.~D.~A. 2014, \mnras, 437, L21,
  \dodoi{10.1093/mnrasl/slt133}

\bibitem[{{D'Ercole} {et~al.}(2008){D'Ercole}, {Vesperini}, {D'Antona},
  {McMillan}, \& {Recchi}}]{dercole08}
{D'Ercole}, A., {Vesperini}, E., {D'Antona}, F., {McMillan}, S. L.~W., \&
  {Recchi}, S. 2008, \mnras, 391, 825, \dodoi{10.1111/j.1365-2966.2008.13915.x}

\bibitem[{{Dobrovolskas} {et~al.}(2014){Dobrovolskas}, {Ku{\v{c}}inskas},
  {Bonifacio}, {Korotin}, {Steffen}, {Sbordone}, {Caffau}, {Ludwig}, {Royer},
  \& {Prakapavi{\v{c}}ius}}]{dobrovolskas2014a}
{Dobrovolskas}, V., {Ku{\v{c}}inskas}, A., {Bonifacio}, P., {et~al.} 2014,
  \aap, 565, A121, \dodoi{10.1051/0004-6361/201322868}

\bibitem[{{Dondoglio} {et~al.}(2022){Dondoglio}, {Milone}, {Renzini},
  {Vesperini}, {Lagioia}, {Marino}, {Bellini}, {Carlos}, {Cordoni}, {Jang},
  {Legnardi}, {Libralato}, {Mohandasan}, {D'Antona}, {Martorano}, {Muratore},
  \& {Tailo}}]{dondoglio2022a}
{Dondoglio}, E., {Milone}, A.~P., {Renzini}, A., {et~al.} 2022, \apj, 927, 207,
  \dodoi{10.3847/1538-4357/ac5046}

\bibitem[{{Dotter} {et~al.}(2008){Dotter}, {Chaboyer}, {Jevremovi{\'c}},
  {Kostov}, {Baron}, \& {Ferguson}}]{dotter2008a}
{Dotter}, A., {Chaboyer}, B., {Jevremovi{\'c}}, D., {et~al.} 2008, \apjs, 178,
  89, \dodoi{10.1086/589654}

\bibitem[{{Gieles} {et~al.}(2018){Gieles}, {Charbonnel}, {Krause},
  {H{\'e}nault-Brunet}, {Agertz}, {Lamers}, {Bastian}, {Gualandris}, {Zocchi},
  \& {Petts}}]{gieles2018a}
{Gieles}, M., {Charbonnel}, C., {Krause}, M. G.~H., {et~al.} 2018, \mnras, 478,
  2461, \dodoi{10.1093/mnras/sty1059}

\bibitem[{{Gratton} {et~al.}(2004){Gratton}, {Sneden}, \&
  {Carretta}}]{gratton04}
{Gratton}, R., {Sneden}, C., \& {Carretta}, E. 2004, \araa, 42, 385,
  \dodoi{10.1146/annurev.astro.42.053102.133945}

\bibitem[{{Katz} \& {Ricotti}(2013)}]{katz13}
{Katz}, H., \& {Ricotti}, M. 2013, \mnras, 432, 3250,
  \dodoi{10.1093/mnras/stt676}

\bibitem[{{Lacchin} {et~al.}(2024){Lacchin}, {Mastrobuono-Battisti}, {Calura},
  {Nipoti}, {Milone}, {Meneghetti}, \& {Vanzella}}]{lacchin2024}
{Lacchin}, E., {Mastrobuono-Battisti}, A., {Calura}, F., {et~al.} 2024, \aap,
  681, A45, \dodoi{10.1051/0004-6361/202347268}

\bibitem[{{Marino} {et~al.}(2019){Marino}, {Milone}, {Renzini}, {D'Antona},
  {Anderson}, {Bedin}, {Bellini}, {Cordoni}, {Lagioia}, {Piotto}, \&
  {Tailo}}]{marino2019a}
{Marino}, A.~F., {Milone}, A.~P., {Renzini}, A., {et~al.} 2019, \mnras, 487,
  3815, \dodoi{10.1093/mnras/stz1415}

\bibitem[{{Marino} {et~al.}(2023){Marino}, {Milone}, {Dondoglio}, {Renzini},
  {Cordoni}, {Jerjen}, {Karakas}, {Lagioia}, {Legnardi}, {McKenzie},
  {Mohandasan}, {Tailo}, {Yong}, \& {Ziliotto}}]{marino2023a}
{Marino}, A.~F., {Milone}, A.~P., {Dondoglio}, E., {et~al.} 2023, \apj, 958,
  31, \dodoi{10.3847/1538-4357/acfca3}

\bibitem[{{Marino} {et~al.}(2024){Marino}, {Milone}, {Legnardi}, {Renzini},
  {Dondoglio}, {Cavecchi}, {Cordoni}, {Dotter}, {Lagioia}, {Ziliotto},
  {Bernizzoni}, {Bortolan}, {Carlos}, {Jang}, {Mohandasan}, {Muratore}, \&
  {Tailo}}]{marino2024}
{Marino}, A.~F., {Milone}, A.~P., {Legnardi}, M.~V., {et~al.} 2024, arXiv
  e-prints, arXiv:2401.06681, \dodoi{10.48550/arXiv.2401.06681}

\bibitem[{{Milone} {et~al.}(2012{\natexlab{a}}){Milone}, {Marino}, {Cassisi},
  {Piotto}, {Bedin}, {Anderson}, {Allard}, {Aparicio}, {Bellini}, {Buonanno},
  {Monelli}, \& {Pietrinferni}}]{milone2012c}
{Milone}, A.~P., {Marino}, A.~F., {Cassisi}, S., {et~al.} 2012{\natexlab{a}},
  \apjl, 754, L34, \dodoi{10.1088/2041-8205/754/2/L34}

\bibitem[{{Milone} {et~al.}(2012{\natexlab{b}}){Milone}, {Marino}, {Cassisi},
  {Piotto}, {Bedin}, {Anderson}, {Allard}, {Aparicio}, {Bellini}, {Buonanno},
  {Monelli}, \& {Pietrinferni}}]{milone2012b}
---. 2012{\natexlab{b}}, \apjl, 754, L34, \dodoi{10.1088/2041-8205/754/2/L34}

\bibitem[{{Milone} {et~al.}(2014){Milone}, {Marino}, {Bedin}, {Piotto},
  {Cassisi}, {Dieball}, {Anderson}, {Jerjen}, {Asplund}, {Bellini}, {Brogaard},
  {Dotter}, {Giersz}, {Heggie}, {Knigge}, {Rich}, {van den Berg}, \&
  {Buonanno}}]{milone2014}
{Milone}, A.~P., {Marino}, A.~F., {Bedin}, L.~R., {et~al.} 2014, \mnras, 439,
  1588, \dodoi{10.1093/mnras/stu030}

\bibitem[{{Milone} {et~al.}(2015){Milone}, {Marino}, {Piotto}, {Renzini},
  {Bedin}, {Anderson}, {Cassisi}, {D'Antona}, {Bellini}, {Jerjen},
  {Pietrinferni}, \& {Ventura}}]{milone2015}
{Milone}, A.~P., {Marino}, A.~F., {Piotto}, G., {et~al.} 2015, \apj, 808, 51,
  \dodoi{10.1088/0004-637X/808/1/51}

\bibitem[{{Milone} {et~al.}(2017{\natexlab{a}}){Milone}, {Piotto}, {Renzini},
  {Marino}, {Bedin}, {Vesperini}, {D'Antona}, {Nardiello}, {Anderson}, {King},
  {Yong}, {Bellini}, {Aparicio}, {Barbuy}, {Brown}, {Cassisi}, {Ortolani},
  {Salaris}, {Sarajedini}, \& {van der Marel}}]{milone2017a}
{Milone}, A.~P., {Piotto}, G., {Renzini}, A., {et~al.} 2017{\natexlab{a}},
  \mnras, 464, 3636, \dodoi{10.1093/mnras/stw2531}

\bibitem[{{Milone} {et~al.}(2017{\natexlab{b}}){Milone}, {Marino}, {Bedin},
  {Anderson}, {Apai}, {Bellini}, {Bergeron}, {Burgasser}, {Dotter}, \&
  {Rees}}]{milone2017b}
{Milone}, A.~P., {Marino}, A.~F., {Bedin}, L.~R., {et~al.} 2017{\natexlab{b}},
  \mnras, 469, 800, \dodoi{10.1093/mnras/stx836}

\bibitem[{{Milone} {et~al.}(2018){Milone}, {Marino}, {Renzini}, {D'Antona},
  {Anderson}, {Barbuy}, {Bedin}, {Bellini}, {Brown}, {Cassisi}, {Cordoni},
  {Lagioia}, {Nardiello}, {Ortolani}, {Piotto}, {Sarajedini}, {Tailo}, {van der
  Marel}, \& {Vesperini}}]{milone2018a}
{Milone}, A.~P., {Marino}, A.~F., {Renzini}, A., {et~al.} 2018, \mnras, 481,
  5098, \dodoi{10.1093/mnras/sty2573}

\bibitem[{{Milone} {et~al.}(2019){Milone}, {Marino}, {Bedin}, {Anderson},
  {Apai}, {Bellini}, {Dieball}, {Salaris}, {Libralato}, {Nardiello},
  {Bergeron}, {Burgasser}, {Rees}, {Rich}, \& {Richer}}]{milone2019}
{Milone}, A.~P., {Marino}, A.~F., {Bedin}, L.~R., {et~al.} 2019, \mnras, 484,
  4046, \dodoi{10.1093/mnras/stz277}

\bibitem[{{Milone} {et~al.}(2023){Milone}, {Marino}, {Dotter}, {Ziliotto},
  {Dondoglio}, {Cordoni}, {Jang}, {Lagioia}, {Legnardi}, {Mohandasan}, {Tailo},
  {Yong}, {Baimukhametova}, \& {Carlos}}]{milone2023a}
{Milone}, A.~P., {Marino}, A.~F., {Dotter}, A., {et~al.} 2023, \mnras, 522,
  2429, \dodoi{10.1093/mnras/stad1041}

\bibitem[{{Partridge} \& {Schwenke}(1997)}]{partridge1997a}
{Partridge}, H., \& {Schwenke}, D.~W. 1997, \jcp, 106, 4618,
  \dodoi{10.1063/1.473987}

\bibitem[{{Pietrinferni} {et~al.}(2021){Pietrinferni}, {Hidalgo}, {Cassisi},
  {Salaris}, {Savino}, {Mucciarelli}, {Verma}, {Silva Aguirre}, {Aparicio}, \&
  {Ferguson}}]{pietrinferni21}
{Pietrinferni}, A., {Hidalgo}, S., {Cassisi}, S., {et~al.} 2021, \apj, 908,
  102, \dodoi{10.3847/1538-4357/abd4d5}

\bibitem[{{Piotto} {et~al.}(2015){Piotto}, {Milone}, {Bedin}, {Anderson},
  {King}, {Marino}, {Nardiello}, {Aparicio}, {Barbuy}, {Bellini}, {Brown},
  {Cassisi}, {Cool}, {Cunial}, {Dalessandro}, {D'Antona}, {Ferraro}, {Hidalgo},
  {Lanzoni}, {Monelli}, {Ortolani}, {Renzini}, {Salaris}, {Sarajedini}, {van
  der Marel}, {Vesperini}, \& {Zoccali}}]{piotto15}
{Piotto}, G., {Milone}, A.~P., {Bedin}, L.~R., {et~al.} 2015, \aj, 149, 91,
  \dodoi{10.1088/0004-6256/149/3/91}

\bibitem[{{Renzini}(2017)}]{renzini17}
{Renzini}, A. 2017, \mnras, 469, L63, \dodoi{10.1093/mnrasl/slx057}

\bibitem[{{Renzini} {et~al.}(2022){Renzini}, {Marino}, \&
  {Milone}}]{renzini2022a}
{Renzini}, A., {Marino}, A.~F., \& {Milone}, A.~P. 2022, \mnras, 513, 2111,
  \dodoi{10.1093/mnras/stac973}

\bibitem[{{Scalco} {et~al.}(2024){Scalco}, {Gerasimov}, {Bedin}, {Vesperini},
  {Nardiello}, {Salaris}, {Burgasser}, {Anderson}, {Libralato}, {Bellini}, \&
  {Rosati}}]{scalco24}
{Scalco}, M., {Gerasimov}, R., {Bedin}, L.~R., {et~al.} 2024, arXiv e-prints,
  arXiv:2403.03262, \dodoi{10.48550/arXiv.2403.03262}

\bibitem[{{VandenBerg}(2023)}]{vandenberg23}
{VandenBerg}, D.~A. 2023, \mnras, 518, 4517, \dodoi{10.1093/mnras/stac3270}

\bibitem[{{Vanzella} {et~al.}(2016){Vanzella}, {de Barros}, {Vasei}, {Alavi},
  {Giavalisco}, {Siana}, {Grazian}, {Hasinger}, {Suh}, {Cappelluti}, {Vito},
  {Amorin}, {Balestra}, {Brusa}, {Calura}, {Castellano}, {Comastri}, {Fontana},
  {Gilli}, {Mignoli}, {Pentericci}, {Vignali}, \& {Zamorani}}]{vanzella16}
{Vanzella}, E., {de Barros}, S., {Vasei}, K., {et~al.} 2016, \apj, 825, 41,
  \dodoi{10.3847/0004-637X/825/1/41}

\bibitem[{{Vanzella} {et~al.}(2020){Vanzella}, {Caminha}, {Calura}, {Cupani},
  {Meneghetti}, {Castellano}, {Rosati}, {Mercurio}, {Sani}, {Grillo}, {Gilli},
  {Mignoli}, {Comastri}, {Nonino}, {Cristiani}, {Giavalisco}, \&
  {Caputi}}]{vanzella20}
{Vanzella}, E., {Caminha}, G.~B., {Calura}, F., {et~al.} 2020, \mnras, 491,
  1093, \dodoi{10.1093/mnras/stz2286}

\bibitem[{{Ventura} \& {D'Antona}(2009)}]{ventura09}
{Ventura}, P., \& {D'Antona}, F. 2009, \aap, 499, 835,
  \dodoi{10.1051/0004-6361/200811139}

\bibitem[{{Ziliotto} {et~al.}(2023){Ziliotto}, {Milone}, {Marino}, {Dotter},
  {Renzini}, {Vesperini}, {Karakas}, {Cordoni}, {Dondoglio}, {Legnardi},
  {Lagioia}, {Mohandasan}, \& {Baimukhametova}}]{ziliotto2023a}
{Ziliotto}, T., {Milone}, A., {Marino}, A.~F., {et~al.} 2023, \apj, 953, 62,
  \dodoi{10.3847/1538-4357/acde76}

\end{thebibliography}
\bibliographystyle{aasjournal}

\end{document}